\documentclass{article}
\usepackage{hyperref}
\pdfoutput=1
\begin{document}

\title{The Life of a Free Soap Film}

\author{H. C. Mayer \& R. Krechetnikov\\
\\\vspace{6pt} Department of Mechanical Engineering
\\ University of California at Santa Barbara}

\maketitle

\begin{abstract}
``The Life of a Free Soap Film'' is a fluid dynamics video for the Gallery of Fluid Motion intended to showcase 
the novel experimental techniques that we have developed in our laboratory to study the retraction of planar 
liquid films of various shapes and sizes as well as other film properties not accessible before. 
Unlike previous experimental studies that have used rupture to initiate 
the retraction of a liquid film from a point, we can produce the uniform release of a soap film
along a straight edge up to many centimeters in length and capture the subsequent retraction of the film and
details concerning the development of the edge using high speed photography. This video only hints at the 
astonishing life of a ``free'' soap films.
\end{abstract}

\section{Motivation}

Although extensively studied in the past, there are still many questions regarding
the retraction of liquid sheets that remain unanswered and many experiments that have remained
unperformed because it had been thought impossible.  This is exemplified by a recent remark in the literature from Nikos 
Savva and John Bush, \emph{``Experimental study of the retraction of a planar film is impractical owing to unavoidable
edge effects and the difficulties inherent in producing a perfectly linear rupture''}\cite{bush}. Prior experimental work has
focused on measurements of the retraction of thin liquid sheets using rupture from a point \cite{ranz}\cite{mysels}\cite{quere}. 
But, is it possible to release a liquid sheet (e.g. a soap film) uniformly along an edge? And, if this can be 
accomplished and extended to all edges supporting the film, what will happen to the ``free'' soap film? Showcased situations in 
this video are a part of our research that highlight the results of the techniques that we have developed in the attempt to
answer such questions.

\section{Experimental Procedures}

The soap films used in our experiments are formed from a solution composed of deionized water, 4\% glycerol, and sodium dodecyl sulfate surfactant at the critical micelle concentration. 
The shape and size of the soap films are set by the nichrome wire frames on which they are formed; the size of the wires ranges between 50 gauge (25 $\mu m$ diameter) and 36 gauge (127 $\mu m$ diameter).  
The wire frames are mounted
to a precision stepper motor which allows the frames to be withdrawn vertically from the solution with a constant specified velocity. 

Once removed from the bath of soap solution, the frame is rotated into a horizontal position to help minimize thickness variations across the film. Soap films produced in this manner are on the order of 1 $\mu m$ thick. The wire frame not only sets the size and shape of the film but also acts as a resistor in a custom-built high voltage circuit connected to a capacitor charged to several thousand volts. When the capacitor is discharged, Joule heating of the wire frame causes the soap film immediately in contact with the frame to rapidly boil thereby detaching the soap film uniformly along the entire frame. The result is a ``free'' soap film and the process of releasing the film and its retraction is captured using a Phantom v5.2 high speed digital camera.  

\section{Video Description}

The video starts with a movie (Soap Film Retraction - Point Rupture) representing the more traditional method of film rupture from a point, in this case caused 
by impacting the film with a cotton swab (such type of rupture can also be initiated by a localized electric spark). The ruptured film retracts outward from a point and the nonuniformity in the shape of the hole is the result of thickness variations in the film.

Our experimental technique allows us to go beyond the classical study of the retraction of liquid films from a point and to capture the retraction of 
planar liquid films detached from their edges. Naturally our first attempts at such experiments involved the uniform release of a soap film along a single edge. An example of such a movie is included in the video (Soap Film Retraction - Single Edge) and shows a portion of a soap film released from a 50 gauge wire along with its retraction. The soap film is seen as the downward moving light gray surface due to the use of reflected light. 

Further movies (Soap Film Retraction - Square Frame, Rectangular Frame, and Circular Frame) show the uniform release of soap films from all edges. We have included examples of square, rectangular, and nearly circular soap films to highlight the variety of shapes and sizes that can be studied. At the start of each movie one can see the initial rapid detachment of the film from the wire frame. While retracting toward the center of the frame at speeds in excess of 1 $m/s$, the films tend to shed droplets - especially at corner locations. At the end of their retraction, droplets of numerous sizes are formed which often collide near the center or travel out of the viewing area. These droplets can be collected and used along with high speed movies for quantitative analysis of the formation of such soap films (e.g. Plateau borders), the influence of the wire frame properties, and of many aspects of retracting films (see Poster 72 - L1 Poster Session - APS DFD 2010). For each frame shape the video includes high speed movies recorded using both transmitted and reflected light. Each type of lighting has a unique benefit: for example transmitted light allows one to capture the process with very short exposure time and this permits more details of the edge of the detached film to be resolved. Droplets formed during the retraction process can also be seen
clearly. On the other hand, reflected light, which represents only a few percent of the incident light used for illumination, results in lower contrast movies, but allows visualization of the surface of the film and can be used to image waves on the soap film.

\end{document}